\documentclass[%
 reprint,
 amsmath,amssymb,
 aps,
]{revtex4-2}
\usepackage{subcaption}
\usepackage{caption}
\usepackage{graphicx}
\usepackage{xcolor}
\usepackage{soul}
\usepackage{dcolumn}
\usepackage{bm}
\usepackage{booktabs}%
\usepackage{graphicx}%
\usepackage{multirow}%
\usepackage{amsmath,amssymb,amsfonts}%
\usepackage{amsthm}%
\usepackage{mathrsfs}%
\usepackage{xcolor}%
\usepackage{float}
\usepackage{seqsplit}

\def\aap{A\&A}

\def\apj{ApJ}
\def\apjl{ApJ}

\def\mnras{MNRAS}

\def\nat{Nature}

\def\prd{Phys.~Rev.~D}

\def\prl{Phys.~Rev.~Lett.}

\begin{document}



\title{JWST evidence for a sharp ``Cosmic Daybreak" at $z=15$} 



\author{Jiashuo (Josh) Zhang$^{1}$}
\author{Tom Broadhurst$^{1,2,3}$}
\email{tom.j.broadhurst@gmail.com}
\author{Tzihong Chiueh$^{5,6}$}
\author{Hsi-Yu Schive$^{5,6}$}
\author{Keiichi Umetsu$^{7}$}
\author{Jose M. Diego$^{10}$}
\author{Paloma Morilla$^{3}$}
\author{Alvaro Pozo$^{1}$}
\author{Chian-Chou Chen$^{7}$}
\author{Jeremy Lim$^{8,9}$}
\author{Pablo G. P\'erez-Gonz\'alez$^{11}$}
\author{Rogier Windhorst$^{12}$}

\affiliation{
$^{1}$\textit{Donostia International Physics Center, DIPC, Basque Country, San Sebasti\'an, 20018, Spain}
$^{2}$\textit{Ikerbasque, Basque Foundation for Science, E-48011 Bilbao, Spain}
$^{3}$\textit{Department of Theoretical Physics, University of the Basque Country UPV/EHU, E-48080 Bilbao, Spain}
$^{5}$\textit{Department of Physics, National Taiwan University, Taipei 10617, Taiwan}
$^{6}$\textit{National Center for Theoretical Sciences, National Taiwan University, Taipei 10617, Taiwan}
$^{7}$\textit{Academia Sinica Institute of Astronomy and Astrophysics (ASIAA), No. 1, Section 4, Roosevelt Road, Taipei 106319, Taiwan}
$^{8}$\textit{Department of Physics, The University of Hong Kong, Hong Kong S.A.R.}
$^{9}$\textit{The Hong Kong Institute for Astronomy and Astrophysics, The University of Hong Kong, Hong Kong S.A.R.}
$^{10}$\textit{Instituto de Física de Cantabria (CSIC-UC), Avda. Los Castros s/n., 39005 Santander, Spain}
$^{11}$\textit{Centro de Astrobiología (CAB), CSIC-INTA, Ctra. de Ajalvir km 4, Torrejón de Ardoz, E-28850, Madrid, Spain}
$^{12}$\textit{School of Earth and Space Exploration, Arizona State University, Tempe, AZ 85287-1404, USA}}









\begin{abstract}
 
 Luminous young galaxies have been uncovered with relative ease by JWST, extending to z=14.5, so it is puzzling that deeper spectroscopy of fainter candidates now finds only interlopers. This redshift `ìmpasse" is underscored by the measured stellar ages of these high-z galaxies, which we show converge to zero by z=15, with a marked absence of earlier star-formation. Taken literally, such a late transition from the Dark Ages to luminous galaxies is unlike the gradual Cosmic Dawn of standard LCDM, but does confirm a key prediction of Wave Dark Matter, $\psi$DM, as a Bose-Einstein condensate. The de Broglie wave pressure resists gravity until a substantial Jeans mass of $4\times 10^9M_\odot$ is overcome at z$=$15, corresponding to a light boson $m_\psi=2.2_{-0.3}^{+0.4} \times 10^{-22}$eV, and similar to independent estimates from lensing anomalies and dwarf galaxies.  Furthermore, the substantial luminosities of the highest redshift galaxies appear to converge to the initial Jeans scale of $\psi$DM, whereas LCDM predictions extend to lower luminosities and larger ages than observed. These contrasting predictions can be definitively tested as JWST observations accumulate, with diametric implications for Dark Matter as heavy particles beyond the Standard Model or ultra-light bosons motivated by the String Axiverse.
 
\end{abstract}

\maketitle

 Many high redshift galaxies discovered by JWST at z$>$10 now have measured ages, spanning $5-150$ Myrs in deep JWST spectroscopy, allowing galaxy formation to be directly charted. High rates of ongoing star formation \cite{Kokorev} are derived within compact galaxies, $r \leq 300$pc \cite{Helton25} with steep, metal-poor UV stellar continua \cite{Bunker2023,Carniani,Donnan_z13.53} and large equivalent width ratios of [C III]/H$\beta$ \cite{Borsani2025,Tang,Scholtz} indicating young ages $\leq 10$Myr. Larger stellar ages of up to $\simeq 150$ Myrs are measured for lower luminosity galaxies, closer to z$=$10, including post-starburst examples \cite{Bradac,Harikane}. Here we compare all reported age estimates at z$>$10 in Figure 1a, revealing the range of ages appears to narrow steadily with increasing redshift, to minimum of $\leq 10$ Myrs by z$=$14, indicating that z$=$15 can be identified empirically as the onset of galaxy formation. We can also use these ages to estimate formation redshifts, $z_f$, by subtracting the stellar age of each galaxy from its cosmological age given by the measured redshift. This produces a flat trend of z$_f$ up to an abrupt limit of z$_f$=15.1, above which there is a complete absence of earlier formed, younger galaxies, significant at $>3\sigma$ level beyond z$>$17.2, as shown in Figure 1b.

\begin{figure*}[ht!]
\centering
\begin{subfigure}{0.43\textwidth}
\centering
\includegraphics[width=\linewidth]{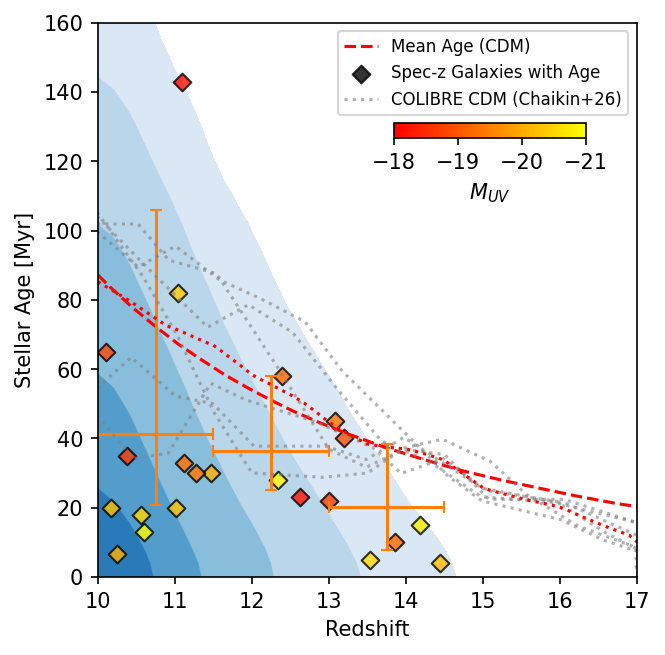}
\caption{}
\end{subfigure}
\hfill
\begin{subfigure}{0.555\textwidth}
\centering
\includegraphics[width=\linewidth]{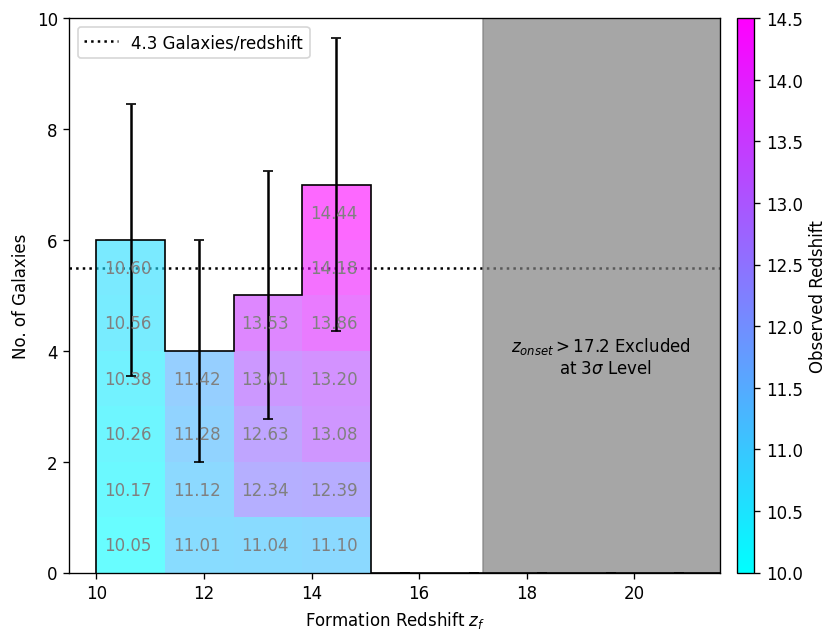}
\caption{}
\end{subfigure}
\caption{\textbf{Left Panel:}
All 22 galaxies with reported stellar ages, compared with their spectroscopic redshifts, indicating by z=15 there is a convergence in age to a minimum of only a few Myrs. The mean observed age and error bars are also indicated. The observations match well the age distribution for $\psi$DM, shaded in blue, for a boson mass of $m_{22}=2$, for which galaxies first form at z=15. The red dashed line shows the LCDM-based prediction for the evolving halo mass function, which has older mean ages and extends to higher redshift than $\psi$DM. This predicted trend is accurately followed by luminous galaxies highlighted by the latest COLIBRE hydro-simulations\cite{Chaikin2026}, shown by the red dotted line, averaged over six individual luminous JWST galaxies (grey dotted lines). and for which star formation extends to $z\simeq 20$. {\bf Right Panel}  Individual formation redshifts calculated for all 22 galaxies after subtracting each reported stellar age from the observed cosmological age given by the spectroscopic redshift (shown on the histogram). A marked onset in formation is evident at z=15.1, above which no earlier star formation is inferred empirically, representing a $3\sigma$ absence of earlier forming galaxies above z$=$17.2. }
\label{Mh_distri}
\end{figure*}

\begin{figure*}[ht!]
\centering
\begin{subfigure}{0.5\textwidth}
\centering
\includegraphics[width=\linewidth]{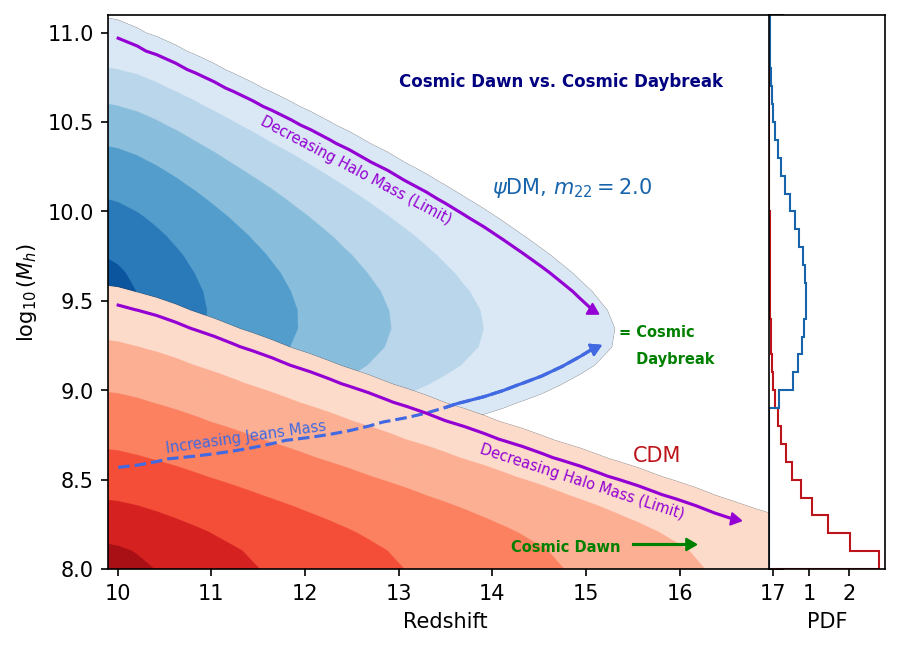}
\caption{}
\end{subfigure}
\hfill
\begin{subfigure}{0.485\textwidth}
\centering
\includegraphics[width=\linewidth]{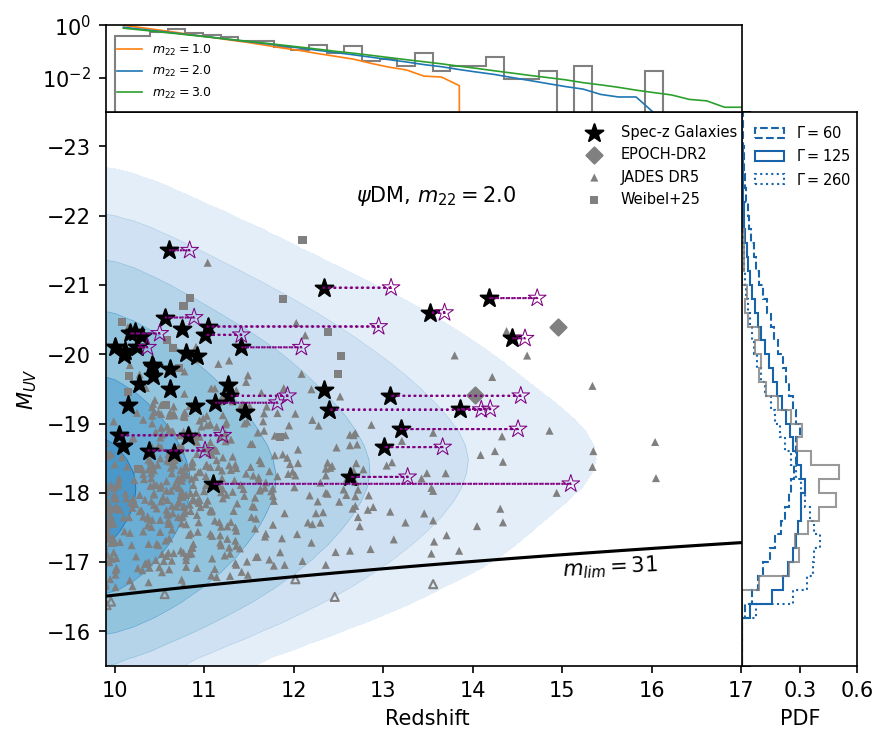}
\caption{}
\end{subfigure}
\caption{\textbf{Left Panel} Comparison of LCDM and $\psi$DM halo mass distributions, showing the convergence to the limiting Jeans mass at z$\sim$15 predicted for $\psi$DM, with a mass of $m_{22}=2$. \textbf{Right Panel} All reported UV Magnitudes vs spectroscopic redshifts (bold points) and also galaxies with photo-z´s reported in the latest deep and wide analyses\cite{Hainline2026, Weibel}, together with two confident photo-z cases ($\beta <-3.0$) marked as grey diamonds\cite{Austin2026}. The observed excess of luminous detections at z$>$13 resembles well the V-shaped prediction for $\psi$DM based on the halo mass distribution shown on the left. While photo-z sample extends to $z\sim16$, purple dotted lines tracing back to when spectroscopically confirmed high-z galaxies have formed do not extend beyond $z\sim 15.1$, as emphasized in Figure 1. }  
\label{scatter_allspecz}
\end{figure*}

The success of JWST in advancing the redshift frontier has relied mainly on F150W dropout selection, for which the interloper rate has not been significant up to the current maximum of z$\simeq $14.5. However, attempts to push further with F200W dropouts appear to have reached a puzzling ``impasse¨, with only $\simeq 20$ viable photo-z candidates identified in the current multi-survey analyses \cite{Pablo,Meyer2024,Castellano,McLeod26,Gandolfi,Austin2026}, for which spectroscopy of the first seven F200W dropouts \cite{Haro,Castellano} finds only interlopers, with z$<$6.6. These cases have bimodal, high and low redshift photo-z solutions, so it can be predicted that spectroscopy of further such F200W dropouts will yield only interlopers with high probability. Hence, this absence above z$=$14.5 is stringent, at least for the high luminosities typical of the galaxies discovered near the limit of z$=$14.44 \cite{Naidu}. Furthermore, the Ly-break at z$=$14.5 causes only a 30\% reduction in F200W flux, so if such luminous galaxies are present at higher redshift they would be easier to recognise by dropping out more strongly in F200W, reaching a maximum depth at z$=$17.2. Hence, it is despite this deepening that nothing has yet been confirmed beyond z=14.5. For the next filter, F277W, corresponding to z$=$18.7-25.5, a complete absence of any dropout candidates is reported by Ref \citep{Castellano}, whereas 3 faint candidates have been proposed as high-z possibilities in the currently deepest F277W pointing of the NGDEEP field, reaching m$_{F277W}=31$, with photo-z´s of z$=$16-20 \citep{Pablo}, for which follow-up spectroscopy is strongly motivated. 
 
If the wider and deeper JWST surveys continue to find the redshift limit only creeps slowly towards z=15 this would be firmly at odds with the long predicted scale-free growth for LCDM, where galaxy formation smoothly extends to lower masses and higher redshifts, well beyond z$=$15, as illustrated in Figures 1a (red dotted line) and Figure 2a (red contours) and by the FIRE-2 and recent COLIBRE hydro-simulations \cite{Sun2023,Chaikin2026}. By contrast, a sudden and late transition to relatively massive galaxy formation is a key prediction of Wave Dark Matter, $\psi$DM, recognized in cosmological simulations \cite{Schive2014a} prior to JWST, where the earliest galaxy appeared at z$=$13, with substantial halo mass of $\sim 10^{9.5}M_\odot$. This onset redshift and initial galaxy mass depends only on the boson mass as the only free parameter for $\psi$DM, which was set to a relatively light value of $m_\psi=0.8 \times 10^{-22}$eV in these first simulations\cite{Schive2014a}. Here, for the first time we derive a scaling of the onset redshift with boson mass using the inherent Jeans condition for $\psi$DM set by the Uncertainty Principle, when self-gravity first overcomes the effective wave pressure on the de Broglie scale, defining an halo Jeans scale $k_J = (16\pi G_N \rho_h)^{1/4} (\frac{m_\psi}{\hbar})^{1/2}$ with $\rho_h$ being the azimuthal mean halo density, below which galaxy formation is strongly suppressed \citep{Huetal2000, Peebles2000,SilkMarsh2014}. For virialized halos, $\rho_h$ follows from the virial over-density $\Delta_v$ for non-relativistic collapse \citep{astro-ph/9710107}, which is also appropriate for $\psi$DM. The Jeans condition says galaxies lighter than, $$M_{min} \approx 1.10 \times10^9 M_\odot(\frac{m_{\psi}}{2.2\times 10^{-22} \text{eV}})^{-3/2} (\frac{1+z}{16})^{3/4} $$
cannot collapse into forming virialized halos. An estimate for the redshift of galaxy formation $z_{onset}$ then follows when $M_{min}$ matches the scale above which heavier halos, corresponding to heavy density peaks, are exponentially suppressed (see Appendix A for details). We confirm in Figure 3 that our analytical $z_{onset}$ estimate (blue dashed curve) agrees with the simulation result of Ref \cite{Schive2014a} within $1\sigma$ uncertainty, indicated by the red star in Figure 3. Using the simulation result as an calibration, giving the red curve, we infer a $\psi$DM mass of $m_\psi \approx 2.18^{+0.41}_{-0.34}\times10^{-22}$eV from the redshift limit of z$_f=$15.1. We comment that above argument is independent of the specific form of $\psi$DM halo mass functions (HMFs), and derives only from physical arguments. We can also estimate $z_{onset}$ by explicitly integrating the $\psi$DM HMFs above halo Jeans limit, as seen by green curve using the same volume as Ref \cite{Schive2014a}, finding good agreement with their simulation result. For the current survey volume of 0.2 sq. degrees FOV achieved by ASTRODEEP \cite{astrodeep} and the latest EPOCHS-DR2 \cite{Austin2026}, a lighter mass of $m_{22} \simeq 1.7$ is inferred from z$_f$=15.1, and by doubling the survey area to 0.4 sq. degrees, the maximum predicted redshift for this $\psi$DM mass increases somewhat to z$_f$=15.4. Regardless the method, the inferred boson masses are in agreement with the range spanned by independent estimates, plotted as horizontal dotted lines in Figure 3 \citep{Pozo, Broadhurst2025, multicopy, dwarfdiffusion}, and will be better constrained as JWST observations accumulate. 

%


 \begin{figure}
     \centering
     \includegraphics[width=\linewidth]{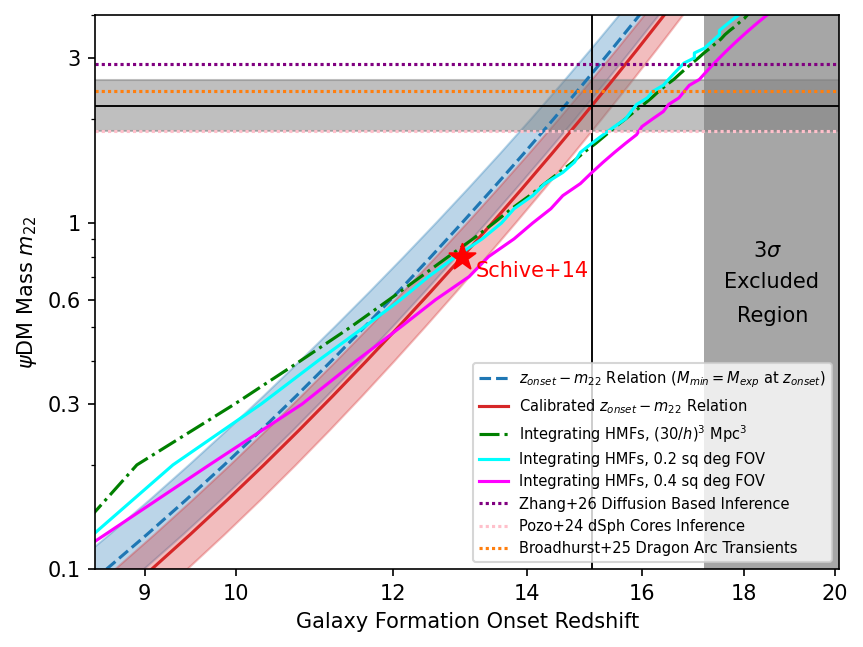}
     \caption{The locus of formation redshift $z_f$ as a function of boson mass (blue dashed curve) obtained through matching the scale of exponentially suppressed density peaks with minimum gravitational collapse scale. The plotted scaling agrees well with the highest redshift halo seen in the cosmological simulation of $\psi$DM at z=13 \citep{Schive2014a}, indicated by the red star, and implying mass of $m_{22}\simeq 2.2$ for z$_f$=15.1 after calibration (red curve). The green solid curve shows the alternative HMF based locus that also agrees with simulation and finds a boson mass of m$_{22} \simeq 1.7$ for z$_f$=15.1. The magenta curve shows the modest increase in maximum redshift we predict for different $\psi$DM masses by doubling the surveyed area. These Jeans and HMF based boson masses are similar to independent estimates from lensing and dynamics shown by horizontal dashed lines \cite{Pozo, Broadhurst2025, multicopy, dwarfdiffusion}. }
     \label{onset_redshift}
 \end{figure}
 
\begin{figure}
    \centering
    \includegraphics[width=\linewidth]{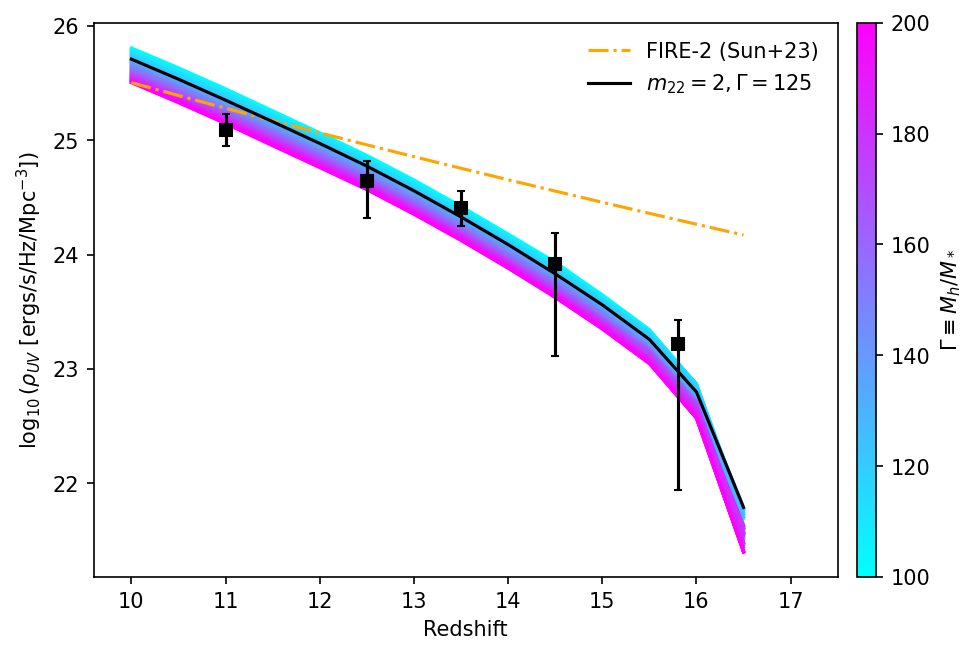}
    \caption{Integrated UV density from the recent multi-survey analysis \cite{Donnan_yet,McLeod26} compared with $\psi$DM ($m_{22}=2.0$) predictions for a range of halo to stellar mass ratio ($\Gamma$) colour coded. It can be seen that despite the simplicity of our model the integrated UV density is well reproduced with $\Gamma = 125$ (black line), consistent with the V-shaped $M_{UV}-z$ distribution of Figure 2b. The dashed orange line is the benchmark FIRE-2 hydro-LCDM modelling \cite{Sun2023}), which is shallower than the data, over predicting the observed decline at high-z.}
    \label{fig:placeholder}
\end{figure}

We now examine the distribution of luminosities for z$>$10, as an additional DM discriminant. The HMF for $\psi$DM is predicted to converge to the initial Jeans mass at z=15 marking the onset of galaxy formation, resulting in a distinctive V-shaped distribution in Figure 2a. This contrasts with the scale free decline with redshift for LCDM, illustrating the slow Cosmic Dawn, also shown in Figure 2a. We compare these contrasting predictions with all reported data shown in Figure 2b, including reliable photo-z´s reported to data in major surveys\cite{Hainline2026,Weibel} and the full spectroscopic data (Table 1). In making our model comparison we assume a fixed halo-to-stellar mass ratio ($\Gamma$) and a constant star formation history for all galaxies, and use \texttt{Bagpipes} \cite{bagpipes} stellar synthesis code to generate the UV absolute magnitude, M$_{UV}$, as a function of age. For more details on the simulated sample, see Appendix. The effect of varying $\Gamma$ is indicated on the right panel of Figure 2b, showing that an increase shifts the prediction of $\psi$DM to higher mean luminosity, with a good fit to the data plotted in Figure 2b achieved with $\Gamma=125$, corresponding to a modest stellar mass of one percent of the halo mass (and a cosmic baryon-star conversion efficiency of $\sim 5\%$). The characteristic V-shaped distribution of $\psi$DM describes well the data, which appears to narrow with increasing redshift, centred on an absolute magnitude of M$_{UV}\sim -19$ by z$\sim$15, well above a flux limit of $m_{lim}=31$, and quite different from the generic prediction of LCDM for which galaxies should be crowded down to the flux limit at any redshift, unlike the data, as can be inferred from Figure 2a.

Selection effects modulate the redshift detection in Figure 2, from broadband selection to the differing depths and filter choices of the major JWST surveys. These have been accurately corrected for in recent measurements of the integrated UV luminosity density\citep{Donnan_yet,McLeod26}, shown in Figure 4 as black data points, allowing for a quantitative comparison with model predictions. For LCDM, the detailed FIRE-2 hydro simulation designed as a benchmark prediction just prior to JWST \cite{Sun2023} is shown in Figure 4, invoking bursty star-formation, matching the lower redshift range but overshoots the latest data at high-z. The relatively steep observed redshift trend is readily matched by our simple  $\psi$DM prediction in Figure 4, for the same mean $\Gamma$ and boson mass used in Figure 2b, assuming only the evolving HMF for $\psi$DM and a simple constant SFH model for galaxies. 


A further distinguishing test can be made with spectroscopy of the lower luminosity 
photo-z candidates at z$>$13, seen in Figure 2b, which if found to have substantial stellar ages of 40Myrs, would support LCDM \cite{Chaikin2026}, as shown in Figure 1a, whereas ages of $<$10Myrs are required by $\psi$DM. This UV spectral slope of the stellar population may also help, as all galaxies close to z$=$15 need to be very young for $\psi$DM as is indeed seen for the highest redshift galaxies, very steep in the UV \cite{Donnan_z13.53,Naidu,Helton25,Wu2025}, including two very blue proposed photo-z candidates $\beta < -3.0$ near z=14.5 \cite{Austin2026}. But for LCDM, UV slopes would be reddened by comparably older stellar ages, or by substantial dust extinction that is predicted in hydro-simulations \cite{Chaikin2026}, providing a further testable prediction motivating deep JWST spectroscopy of lower luminosity candidates at z$>$13.5.

It has been over a year since the z=14.44 record redshift was established, despite the high luminosities of the highest redshift galaxies. This also appears to be the case for the large lensing programs underway with JWST, where 
over 100 massive lensing clusters reach highly magnified sources, but nothing has been found beyond z=15 despite the positive magnification biassed faint-end luminosity function for LCDM (see Appendix). Hence, a continued absence of higher redshift galaxies may soon falsify LCDM and by the same token may bolster $\psi$DM. Galaxy formation is also delayed by Warm Dark Matter (WDM), as free streaming smooths away substructure but may be distinguished by lensing from the ubiquitous granular substructure of $\psi$DM\cite{Amruth,Broadhurst2025}. Other distinguishing measurements include filament induced elongation of young galaxies \cite{Pandaya,Pozo_nat} and stellar dynamics heated by internal de Broglie wave action\cite{Chen1,Pozo}. Currently for $\psi$DM the boson mass favoured here of $\simeq 2\times 10^{-22}$eV is consistent with the lensing and dynamical estimates, but substantial improvements in observational precision of these estimates are feasible thanks to JWST, providing stringent tests of these leading DM contenders.

\begin{acknowledgments}

\noindent TJB thanks Richard Ellis and Guido Roberts-Borsani for inspiring discussions. J.Z. and TJB. acknowledge the CEX2024-001491-S grant, funded by MICIU/AEI/10.13039/501100011033. TJB. is supported by the Spanish Grant PID2023-149016NB-I00 (MINECO/AEI/FEDER, UE). J.L. acknowledges the RGC/GRF 17304425 grant. K.U. acknowledges support from the National Science and Technology Council of Taiwan (grants NSTC 112-2112-M-001-027-MY3 and NSTC 115-2112-M-001-027-) and the Academia Sinica Investigator Award (grant AS-IA-112-M04). H.Y.S. is supported by the National Science and Technology Council (NSTC) of Taiwan under Grant No NSTC 115-2628-M-002-019-MY4. Also PGP-G acknowledges support from grants PID2022-139567NB-I00 and PID2025-169195NB-I00 funded by Spanish Ministerio de Ciencia, Innovaci\'on y Universidades MICIU/AEI/10.13039/501100011033, and the European Union FEDER program {\it Una manera de hacer Europa}. 

\end{acknowledgments}



\appendix
\section{$\psi$DM $z_{onset}-m_{22}$ scaling}

The lowest mass galaxies formed in $\psi$DM at a given redshift correspond to the minimum mass whose self-gravity can over-come the effective wave pressure on the de Broglie scale, within which the Uncertainty Principle prevents confinement of material. A corresponding Jeans scale can be defined, setting the lower halo mass limit, such that \citep{Huetal2000}:
\[ k_J = (16\pi G_N \rho_h)^{1/4} \bigg{(}\frac{m_\psi}{\hbar}\bigg{)}^{1/2}, \] 
\noindent where $\rho_h$ is mean halo density. For virialized halos, $\rho_h$ is simply given by $\Delta_v \overline{\rho_m}$ with $\Delta_v$ the virial over-density constant obtainable from fitting to $\Lambda$CDM simulation \citep{astro-ph/9710107}:

\[ \Delta_v = \frac{18\pi^2+82(\Omega_m-1)-39(\Omega_m-1)^2}{\Omega_m(z)},\] 
\[ \Omega_m(z) = \Omega_{m0}(1+z)^3 \bigg{(} \frac{H_0}{H(z)}\bigg{)}^2, \]

\noindent The above Jean scale $k_{J}$ corresponds to a minimum halo mass scale of:

\begin{align*}
    &M_{min} = \frac{4\pi}{3} \Delta_v \overline{\rho_m} \bigg{(} \frac{2\pi}{k_J} \bigg{)}^3 =\\ &(1+z)^{3/4} \Delta_v(z)^{1/4} m_{22}^{-3/2} \bigg{(} \frac{\Omega_{m,0}h^2}{0.143}\bigg{)}^{1/4} 1.23\times10^8 M_\odot,
\end{align*} 
\noindent where $m_{22} \equiv m_\psi/10^{-22}$, and we adopt $\Omega_{m,0} = 0.315, \; H_0 = 67.37 $ km/s/Mpc as measured by \citet{Planck2018}. Note this Jeans cut-off is lower than $M_{1/2}$ characterizing where $\psi$DM linear matter power spectrum (and HMF) is halved compared to that of LCDM\cite{Schive2016}, because $M_{1/2}$ encodes earlier suppression from radiation-dominated epoch, whereas $M_{min}$ prohibits the collapse of small over-densities in matter-dominated epoch. In Figure 5, we plot $M_{min}$ at different redshifts for $m_{22}=1,3,9,27$ as coloured dashed curves, where it can be seen the lighter the boson mass, the heavier the minimum halo mass. 

Very massive galaxies at high redshifts are also suppressed, owing to exponentially suppressed abundance of large over-densities. Hence, we can arrive at the corresponding maximum, or onset, redshift of galaxy formation for each $m_{22}$ by equating $M_{min}$ with the turnover scale of the HMFs. For this purpose we adopt python package \texttt{hmfmake} \cite{1306.6721} to construct the Sheth-Tormen HMFs as a function of redshift, and measure the corresponding exponential cut-off mass scales $M_{exp}$ through fitting Sheth-Tormen HMFs with a Press-Schechter form $ \frac{dn}{d\ln M} = A M^\alpha e^{-(\frac{M}{M_{exp}})^\beta}$. In Figure 5, this fitted $M_{exp}$ at different redshift is plotted as a black solid line, with the numerical uncertainty in fitting shown as gray shaded region. This figure then provides pictorial representation for $z_{onset}$ as a function of $\psi$DM mass, corresponding to where intersection of curves occurs as shown. $z_{onset}-m_{22}$ scaling thus obtained, along with uncertainty, is shown in Figure 3 as the blue dashed line.

 \begin{figure}
     \centering
     \includegraphics[width=\linewidth]{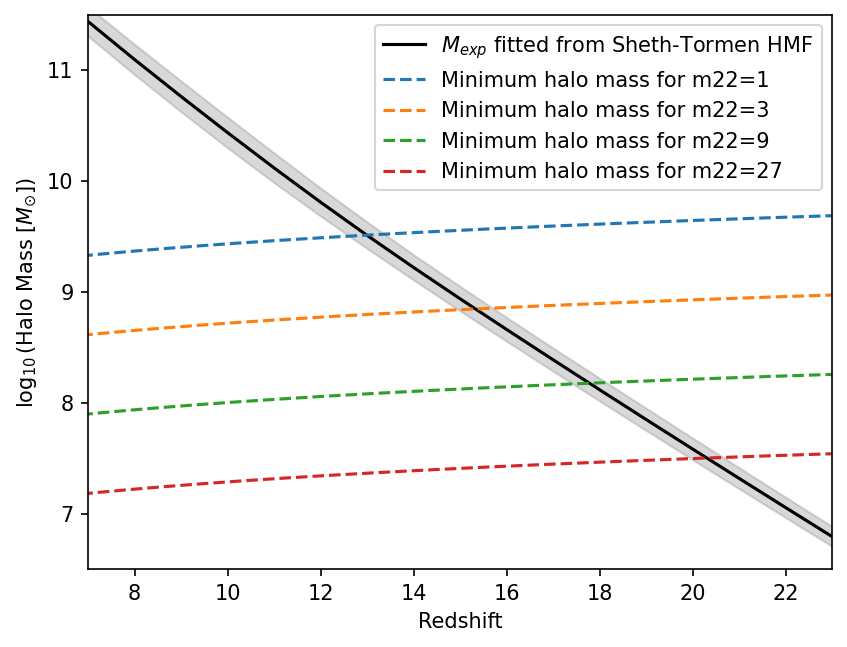}
     \caption{Lower cut-off (i.e. Jeans limit) mass $M_{min}$ for four choices of boson masses of $m_\psi=1,3,9,27 \times 10^{-22}$eV, dashed curves, compared with the exponential cut-off mass scale $M_\ast$ (solid black curve) estimated for the Sheth-Tormen HMFs. Shaded region indicates the numerical fitting uncertainty. An onset redshift can be defined as where $M_{min}$ and $M_{exp}$ meet, as shown in Figure 3.}
     \label{minimumhalomasss}
 \end{figure}

\section{simulated sample}

In Figure 2b, we compare all reported spec-z galaxies at $z>10$ from multiple JWST field surveys and the lensed examples, with our $\psi$DM based predictions for the distribution of luminosity with redshift. The corresponding simulated sample was obtained through numerical sampling of HMFs, where we started from an initial redshift of 30, and 'track' the formation of galaxies in small redshift increments of 0.05, with the number of sampled galaxies obtained through integrating Sheth-Tormen HMFs with simulation-based $\psi$DM low-mass end suppression\citep{Schive2016} included. As quantum pressure forbids over-densities on a mass scale below $M_{min}$ to collapse, our HMF integration limits are from $M_{min}(z, m_{22})$ to $10^{17} M_\odot$ for $\psi$DM model. For LCDM model without low-mass end suppression on HMF, for comparison done in Figure 2a, we adopt integration limits from $10^{8} M_\odot$ to $10^{17} M_\odot$ instead. We assume a fiducial area of 0.2 square degree matching the large area achieved by ASTRODEEP catalog\cite{astrodeep} and latest EPOCHS-DR2 \cite{Austin2026}. 

At each redshift, halo masses are randomly drawn from the underlying HMF and ages are randomly assigned from the cumulative age distribution, measured from all previously injected galaxies with ages evolved to the current working redshift. Figure 1 presents the stellar age distribution thus obtained for $\psi$DM, showing an increased upper limits at lower redshift and convergence to zero age near $z_{onset}\sim 15$, while the red dotted line showing mean stellar age at different redshifts for LCDM extends to higher redshifts. The stellar masses of sampled galaxies are assigned with a fixed halo-to-stellar mass ratio, $\Gamma\equiv M_h/M_*\simeq 125$. Then the stellar masses, along with respective assigned ages, are converted into $M_{UV}$ assuming a constant SFH for galaxies. More specifically, we used public-available python library \texttt{bagpipes} with default BC03 stellar grids\footnote{Changing to BPASS v2.2.1 grids with same SFH model would make young galaxies $\lesssim$ 20 Myr brighter by maximally $\lesssim 0.45$ mag, and hence extends the brighter end of $M_{UV}$ distribution. The overall distribution is same for both $\psi$DM and LCDM models and does not alter our discussion. } to obtain Age-$M_{UV}$ curve, fixing to $M_*=10^9M_\odot$ and assuming observationally motivated metallicity $Z=0.1Z_\odot$ and ionization parameter $\log U = -2$ for high spec-z galaxies \citep{Tang, Borsani2025}. The final luminosity-redshift distribution of simulated galaxies is presented by blue contours in Figure 2b.

\section{Anticipated Number of Highly Magnified Galaxies/Arcs}

The fact that there has been no detection of strongly lensed high-z galaxies beyond redshift 15 is startling for LCDM given the large number ($\sim 150$ through a rough estimate) of clusters surveyed, as we show below with a quick estimate on the number of highly magnified $15\leq z\leq16$ galaxies anticipated. Magnification bias \cite{Tom1995, Leung2018, Zhang_interloper} finds the density of lensed galaxies to be

\[ n_{len} (z,\mu_z,m_{lim}) = \frac{\int^{m_{lim}+2.5\log_{10}\mu_z}_{-\infty} \phi(m',z)dm'}{\mu_z}, \]

\noindent where $\phi(m)$ is the underlying LF, $m_{lim}$ is the magnitude limit, and $\mu_z$ the lensing magnification factor whose appearance in the bottom reflects that the cosmic volume is simultaneously reduced as the detection limit is lowered (by $2.5\log_{10}\mu$). Above expression converts to an estimate of the total number of lensed $15\leq z\leq16$ galaxies through explicit integration over redshift, given the knowledge of underlying LF and also covered areas by differently magnified regions. For highly magnified regions with $\mu>15$, Ref
\cite{Acebron2020} (Figure 3) demonstrates that different lensing clusters share the same scaling law for the covered areas above a given magnified factor, with an median relation of $A(>\mu) \approx 10\text{[sq. arcmin]}/\mu$ that we shall adopt. We may further differentiate above scaling to give $S(\mu) \approx 10/\mu^2$, such that $A(>\mu) = \int_\mu^{\infty} S(\mu) d\mu = 0-(-10/\mu) = 10/\mu $, with which, we calculate the total number of lensed arc (or strongly lensed galaxies with $\mu >\mu_0$) above detection threshold $m_{lim}$ as: 

\[ N = N_{cluster}  \int dz \int_{\mu_0} d\mu_z \;  n_{len}(z,  \mu_z, m_{lim}) S(\mu_z) \frac{dV}{d\Omega dz}\bigg{|}_z \]

\noindent where $N_{cluster}$ is number of clusters currently imaged by JWST $\approx 150$, and $\frac{dV}{d\Omega dz}$ is incremental cosmic volume per square arcmin. Using FIRE-2 LFs and $m_{lim}=31$, we anticipate $\sim19$ $15\leq z\leq 16$ galaxies should have been imaged already given the large sample of lensing clusters that have been imaged today. 

While above calculation is based on a specific (FIRE-2) UV LFs model, we emphasize that the detection of high-z galaxies are made easier with gravitational lensing for LCDM in general, as the scale free structure formation suggests an ever increasing abundance of fainter galaxies, which compensates for the $1/\mu$ reduction in cosmic volume and leads to $n_{len}>n_{unlensed}$, i.e. positive magnification bias. Those fainter $z>15$ galaxies are non-existent in $\psi$DM, hence gravitational lensing would only be magnifying total emptiness, as is hinted by the current no detections of highly magnified galaxies.  

\begin{table*}[h]
    \centering
    \begin{tabular}{c|c|c|c|c|c|c|c}
        Galaxy & $z_{spec}$ & $\mu$ & $M_{UV}$ &  Age [Myr] & $z_{form}$ & $\log_{10}(M_*[M_\odot])$ &  References \\
        \hline
        MoM-z14 & 14.44 & - & $-20.23\pm 0.06$ & 4$^a$ &14.59 & 8.1 & \cite{Naidu} \\ 
        GS-z14-0 & 14.18 & - & $-20.81\pm 0.16$ & 15 & 14.72 & 8.72 &  \cite{Helton}\\ 
        GS-z14-1 & 13.86 & - & $-19.21\pm 0.15$ & 10$^b$ & 14.20 & 7.57 & \cite{Wu2025} \\  
        PAN-z14-1 & 13.53 & - & $-20.6\pm 0.2$ & 5 & 13.69 & 8.23 & \cite{Donnan_z13.53} \\ 
        GS-z13-0 & 13.20 & - & $-18.92\pm 0.05$ & 40 & 14.51 & 7.7 & \cite{Hainline24}\\ 
        UNCOVER-13077 & 13.079 & 2.3 & $-19.4 \pm 1.8$ & 45 & 14.54 & 7.92 & \cite{WBJ23}\\ 
        JADES-GS-z13-1-LA & 13.006 & - & $-18.66\pm 0.04$ & 22 & 13.67 & 7.74 & \cite{Witstok2025} \\ 
        GS-z12-0 & 12.63 & - & $-18.23\pm 0.16$ & 23 & 13.27 & 7.64 & \cite{CL2023}\\
        UNCOVER-38766 & 12.393 & 1.5 & $-19.2 \pm 0.5$ & 58 & 14.10 & 8.63 & \cite{WBJ23}\\ 
        CAPERS-EGS-65480 & 12.344 & - & $-19.48 \pm 0.26$ & - & - & - & \cite{Borsani2025} \\
        GHZ2/GLASS-z12 & 12.338 & 1.3 & $-20.96 \pm 0.05$ & 28 & 13.09 & 9.03 & \cite{Zavala2025}\\ 
        EGS-z11-R0 & 11.452 & - & $-19.16 \pm 0.17 $ & - & - & 9.64 & \cite{Rodighiero26, Kreilgaard26} \\
        CEERS-1 & 11.416 & - & $-20.1\pm 0.1$ & 30  & 12.09 & 8.3 & \cite{Haro} \\
        GS-z11-1 & 11.280 & - &  -19.4 $\pm$ 0.05 & $<30$ & $<11.96$ & 8.03 & \cite{Scholtz}\\
        JADES-GS-20015720 & 11.267 & - & $-19.56 \pm 0.17$ & - & - & - & \cite{Borsani2025, Tang}\\ 
        GS-z11-0-A & 11.122 & - & $-19.30 \pm 0.05$ & 33 & 11.82 & 8.43 & \cite{Witstok26} \\  
        BulletArc-z11 & 11.100 & 14.0 & $-18.13\pm 0.28$ & 143 & 15.10 & 8.2 & \cite{Bradac} \\ 
        CEERS2-588 & 11.040 & - & $-20.4\pm 0.10$ & 82 & 12.95 & 9.05 & \cite{Harikane}\\ 
        CAPERS-UDS-z11 & 11.013 & - & $-20.28 \pm 0.08 $ & $<20$ & $<11.45$ &  8.7 & \cite{Kokorev} \\ 
        MoM-z11-1 & 10.921 & - & $-19.98 \pm 0.24$ & - & - & - & \cite{Borsani2025} \\ 
        JADES-GS-20177294 & 10.893 & - & $-19.25\pm 0.09$ & - & - & - & \cite{Borsani2025,Tang} \\ 
        CAPERS-EGS-43539 & 10.816 & - & $-18.82\pm 0.36$ & - & - & - & \cite{Borsani2025} \\ 
        MoM-z11-2 & 10.803 & - & $-20.02\pm 0.22$ & - & - & - & \cite{Borsani2025} \\
        EGS-22637 & 10.750 & - &  $-20.36\pm 0.09$ & - & - & - & \cite{Borsani2025, Pollock26} \\
        GHZ4 & 10.660 & 1.64 & $-18.58 \pm 0.11$ & - & - & - & \cite{Napolitano25}\\ 
        JADES-GS-20176151 & 10.619 & - & $-19.50\pm 0.14$ & - & - & - & \cite{Borsani2025,Tang}\\ 
        EGS-69 & 10.619 & - & $-19.78 \pm 0.21$ & - &  - & - & \cite{Borsani2025} \\
        GN-z11 & 10.603 & - & $-21.50\pm 0.02$ & 13 & 10.84 & 8.73 & \cite{Bunker2023} \\ 
        CAPERS-UDS-z10 & 10.562 & - & $-20.53 \pm 0.09$ & $<18$ & $<10.89$ & 8.3 & \cite{Kokorev} \\
        GHZ7 & 10.430 & 1.2 & $-19.82 \pm 0.05$ & - & - & - & \cite{Napolitano25}\\ 
        GS-20030902 & 10.425 & - & $-19.68 \pm 0.09$ & - & - & - & \cite{Borsani2025}\\ 
        GS-72355 & 10.422 & - & $-19.85 \pm 0.07$ & - & - & - & \cite{Borsani2025,Weibel} \\
        GS-z10-0 & 10.380 & - & $-18.61 \pm 0.1$ & 35 & 11.01 &  7.58 & \cite{CL2023} \\ 
        UDS-52799 & 10.302 & - & $-20.26\pm 0.08$ & - & - & - & \cite{Borsani2025,Weibel} \\
        CAPERS-COS-109917 & 10.272 & - & $-19.57 \pm 0.11$ & - & - & - & \cite{Borsani2025,Tang} \\ 
        UNCOVER-37126 & 10.255 & 2.19 & $-20.10 \pm 0.05$ & 6.8 & 10.13 & 7.77 & \cite{MChaves26}\\  
        GHZ8 & 10.231 & 1.2 & $-20.32 \pm 0.07$ & - & - & - & \cite{Napolitano25} \\ 
        JD & 10.170 & - & $-20.3\pm 0.2$ & 20 & 10.50 & 8.1 & \cite{Hsiao2024} \\ 
        GHZ9 & 10.145 & 1.4 & $-19.27 \pm 0.04$ & - & - & 8.69 & \cite{Napolitano25, Nap25_GHZ9}\\ 
        MoM-z10-1 & 10.116 & - & $-20.05\pm 0.12$ & - & - & - & \cite{Borsani2025}\\
        CEERS-64 & 10.100 & - & $-19.99 \pm 0.09$ & - &- & - & \cite{Haro, Borsani2025} \\ 
        GLASS-z11-17225 & 10.090 & 1.55 & $-18.68\pm 0.13$ & - & - & - &  \cite{Napolitano25} \\
        UNCOVER-26185 & 10.054 & 3.9 & $-18.83 \pm 0.07$ & 65 & 11.22 & 8.23& \cite{AMarquez26} \\ 
        CEERS-80041 & 10.010 & - & $-20.1\pm 0.10$ & - & - &  9.1 &  \cite{AHaro2023}\\
        
            \end{tabular}
    \caption{All reported galaxies at $z>10$ from JWST surveys to date, including reported stellar age estimates and magnification, $\mu$, for lensed sources. The formation redshift is inferred from the spectroscopic redshift minus the stellar age. \newline
    $^a$: this is the age by which MoM-z14 have formed 50\% of its stellar mass. \newline 
    $^b$: this is the age by which GS-z14-1 have formed 67\% of its stellar mass.   }
    \label{tab:speczsystems}
\end{table*}

\bibliography{reference}

\end{document}